\documentstyle[psfig,aps,prl,amssymb]{revtex}
\begin{document}
\title{$SU(N) \times S_{m}$-Invariant Eigenspaces of $N^{m} \times N^{m}$
Mean Density Matrices}

\author{Paul B. Slater}
\address{ISBER, University of
California, Santa Barbara, CA 93106-2150\\
e-mail: slater@itp.ucsb.edu,
FAX: (805) 893-7995}

\date{\today}

\draft
\maketitle
\vskip -0.1cm

\begin{abstract}
We extend to additional  probability measures
and scenarios, certain of the recent results of Krattenthaler and Slater 
(quant-ph/9612043) ---  whose
 original motivation was to obtain quantum analogs of
seminal work on universal data compression of Clarke and Barron. 
KS obtained explicit formulas for the 
eigenvalues
and eigenvectors of the $2^{m} \times 2^{m}$ density matrices derived  by
averaging the $m$-fold tensor products with themselves of the $2 \times 2$
density matrices. The weighting was done with respect to a one-parameter
($-\infty < u < 1$)
family of probability distributions, all the members of 
which are spherically-symmetric 
($SU(2)$-invariant) over the ``Bloch
sphere'' of two-level quantum
 systems. For $u = {1 \over 2}$, one obtains the normalized volume
element of the minimal monotone (Bures) metric.
In this paper, analyses parallel to those of KS are
conducted, based on an alternative
``natural'' measure on the density matrices recently proposed by \.Zyczkowski,
Horodecki, Sanpera, and Lewenstein  (quant-ph/9804024).
The approaches of KS and that based on ZHSL are found to yield
$\lfloor 1 + {m \over 2} \rfloor$
identical $SU(2) \times S_{m}$-invariant
 eigenspaces (but not coincident eigenvalues for $m > 3$).
Companion results, based
on the $SU(3)$ form of the ZHSL measure, are presented for the
twofold and threefold tensor products of the $3 \times 3$ density matrices.
In the former case, there are six invariant eigenspaces of 
nine-dimensional Hilbert space, and in the latter,
seventeen, of twenty-seven dimensional Hilbert space.
We find a rather remarkable limiting procedure (selection rule)
 for recovering from these
analyses,  the 
(permutationally-symmetrized)
 multiplets  of $SU(3)$
 constructed from two or three particles. We also analyze  the scenarios
(all for $m=2$) $N=4$,
$N=2 \times 3$, $N=2 \times 3 \times 2$ and $N= 3 \times 2 \times 2$
and, in addition,  generalize the ZHSL measure, so that it incorporates a 
 family of (symmetric) Dirichlet distributions --- rather than
simply the uniform distribution  --- defined on the $(N-1)$-dimensional
simplex of eigenvalues.
\end{abstract}


\vspace{-0.1cm}

\newpage

\tableofcontents

\section{INTRODUCTION}

Wootters, in a paper entitled ``Random Quantum States,'' stated that ``there
does not seem to be any natural measure on the set of all mixed
states'' \cite{wootters}.
Relatedly, Jones \cite{jones} asserted that the problem in the
Bayesian treatment of ``the more realistic experimental case of mixed input
states \ldots lies in selecting a good prior on mixed density matrices''.
Also, somewhat earlier, Band and Park \cite{band}  were concerned with
 the ``lack of a rational axiom
of prior distribution over the entire domain of density operators''.

Contrastingly, however, \.Zyczkowski, Horodecki, Sanpera and Lewenstein
\cite{zycz} have recently proposed ``a natural measure in the space of
density matrices'' and used it to ``estimate the volume of separable states,
providing numerical evidence that it decreases exponentially with the
dimension of the composite system''. 
Also, the present author, in a series of papers
 \cite{slat1,slat2,slat3,slat4,slat5} (cf. \cite[secs. 4 and 5]{hall} and
\cite{hall2})
 has pursued the strategy of attempting to normalize the
volume elements of certain ``monotone'' 
Riemannian metrics \cite{petzsudar} --- in
particular, the minimal (Bures \cite{hub,braun2})
 and maximal monotone ones --- and use the
results for information-theoretic/Bayesian, as well as thermodynamic
 purposes.
In \cite{slat4}, it was  concluded --- supportive of certain earlier results
of Petz and Toth \cite{petztoth} --- that
 among the family of monotone metrics,
the maximal was most {\it noninformative} in character.

In this communication, we seek to compare and contrast these various
measures based on monotone metrics with the measures newly
 presented in \cite{zycz}.
In doing this, we immediately encounter a most interesting difference.
While in \cite{slat1,slat2,slat3,slat4,slat5}, the measures are
directly defined over the $(N^2-1)$-dimensional convex sets of 
$N \times N$ density matrices, in \cite{zycz} they are taken to be
the products of the ($N^2$-dimensional) Haar 
(invariant) measure over $U(N)$
and the uniform measure over the $(N-1)$-dimensional simplex spanned by
the $N$ (nonnegative) eigenvalues of
 the $N \times N$ density matrices. Thus, in
\cite{zycz}, the  measures are defined in the (larger) $(N^2+N-1)$-dimensional
space.
\section{THE CASE  $N=2$}
\subsection{The subcase $m=1$}
We have been unable
to obtain an
explicit (degenerate/singular)
 transformation or mapping to pass from these higher-dimensional  measures
 to the lower-dimensional ones, even for the case $N=2$.
Notwithstanding this, we still proceed analytically by, first,
 averaging the $2 \times 2$ density matrices with respect to
these differing measures. Such matrices are parameterizable, using 
the conventional ``Bloch sphere'' representation \cite{braunstein}, as,
\begin{equation} \label{densitymatrix}
{1 \over 2} \pmatrix{ 1 + r \cos{\vartheta} & r \sin{\vartheta}
 \cos{\vartheta} +
\mbox{i}
r \sin{\vartheta} \sin{\phi} \cr
r \sin{\vartheta} \cos{\vartheta} - \mbox{i} r
 \sin{\vartheta} \cos{\vartheta} & 1 - r \cos{\vartheta} \cr}
\end{equation}
($0 \leq r \leq 1$,\quad $0 \leq \vartheta  \leq \pi$,
\quad $0 \leq \phi < 2 \pi$). Averaging these matrices with respect to
 any of the (normalized) measures used by Slater, expressed in the 
same variables, we
 simply obtain, as seems evident, the familiar density matrix
of the {\it fully} mixed state,
\begin{equation} \label{mixed1}
\pmatrix{1/2 & 0\cr
         0 & 1/2\cr}.
\end{equation}
We obtain the  same result using the measure
 of \.Zyczkowski {\it et
 al} \cite{zycz}, but now expressing (\ref{densitymatrix})
not
 in terms of $r, \vartheta$ and $\phi$, but rather, say,
spherical polar coordinates in {\it four}-dimensions
 \cite[eq. (3.129)]{biedenharn}. We, then, perform the
 averaging making use of the associated
Haar measure \cite[eqs. (3.130), (3.131)]{biedenharn}.
\subsection{The subcases $m=2,3$}
We enter a less intuitive realm apparently, however, if
we average 
with respect to  the same normalized measures (that is,
probability distributions),
 the $4 \times 4$ density matrices, which are
 the tensor products of the $2 \times 2$ density matrices
with themselves.
(We note that these averaged matrices and all the ones subsequently
discussed, describe by their very construction,
states which  are {\it separable}, that is, {\it classically correlated}
\cite{werner}.
For an interesting discussion of the significance of the tensor product
in {\it quantum} computation, see \cite{jozsaqc}.)
In fact, in \cite{kratt} (cf. \cite{slatbures}),
 Krattenthaler and Slater derived a general
formula for the entries of $2^m \times 2^m$ density matrices of this type
(as well as their eigenvalues and eigenvectors),
when the averaging was performed with respect to any member of the 
one-parameter ($u$) family
of probability distributions over the Bloch sphere,
\begin{equation} \label{prob}
{\Gamma({5 \over 2} - u) r^2 \sin{\vartheta}
 \over \pi^{{3 \over 2}} \Gamma(1-u) (1-r^2)^u}. \qquad -\infty < u < 1
\end{equation}
The {\it asymptotics} [$m \rightarrow \infty$] of the {\it relative entropy} of
the $2^m \times 2^m$ $m$-fold tensor products with respect to their 
corresponding averages,
as a function of $u$,
was also obtained in \cite{kratt}, and its relevance to
``universal quantum coding'' discussed (cf. \cite{slatsqueezed,jozsa}).

The case $u = {1 \over 2}$  of (\ref{prob})
 gives us the (normalized) volume element of
the minimal (Bures) monotone metric (cf. \cite[eq. (30)]{hall}).
The maximal monotone metric corresponds to $u = {3 \over 2}$. 
Its volume element is not normalizable over the {\it entire} Bloch
sphere. But it is normalizable, if we remove an
 $\epsilon$-neighborhood of the 
spherical boundary ($r=1$). Then, it is possible --- proceeding in Cartesian
coordinates and taking the limit
($\epsilon \rightarrow 0$) of a certain ratio --- to obtain
{\it two}- and {\it one}-dimensional (marginal) probability
 distributions \cite{slat5}.
 (The one-parameter family (\ref{prob}) was employed in \cite{kratt}
for its computational tractability, in addition to its intrinsic interest.
This line of work constituted an effort to develop quantum analogs
of recent seminal results of Clarke and Barron
concerning universal data compression
\cite{clarke1,clarke2}.)

Now, let us, for the scenarios $m=2$ and $3$,
 average the $m$-fold tensor products with
themselves of the 
 $2 \times 2$ density matrices, first,
with respect to the (normalized) measure utilized by \.Zyczkowski {\it et al},
which is, representing the Euler-Rodrigues parameters
in terms of spherical polar coordinates in four
dimensions ($ 0 \leq \phi < 2 \pi,0 \leq \theta 
\leq \pi,0 \leq \chi \leq \pi$),
 \cite[eqs. (3.129)-(3.131)]{biedenharn},
proportional to
\begin{equation} \label{prob2}
{{\sin^{2}{\chi}}} \sin{\theta}.
\end{equation}
The $4 \times 4$ and $8 \times 8$ results obtained are precisely the same, as
those found by averaging the
$2^m \times 2^m$ density matrices (parameterized as in (\ref{densitymatrix}))
 using 
the probability distribution for $u=-2$ in (\ref{prob}).
For $m=2$, this common average is
\begin{equation} \label{resultm2}
\pmatrix{{5 \over 18} & 0 & 0 & 0 \cr
          0 & {2 \over 9} & {1 \over 18} & 0 \cr
          0 & {1 \over 18} & {2 \over 9} & 0 \cr
          0 & 0 & 0 & {5 \over 18} \cr},
\end{equation}
(not simply the diagonal matrix with entries ${1 \over 4}$,
as might have been naively anticipated on the basis of (\ref{mixed1}))
and for $m = 3$,
\begin{equation} \label{resultm3}
\pmatrix{{1 \over 6} & 0 & 0 & 0 & 0 & 0 & 0 & 0 \cr
    0 & {1 \over 9} & {1 \over 36} & 0 & {1 \over 36} & 0 & 0 & 0 \cr
    0 & {1 \over 36} & {1 \over 9} & 0 & {1 \over 36} & 0 & 0 & 0 \cr
    0 & 0 & 0 & {1 \over 9} & 0 & {1 \over 36} & {1 \over 36} & 0 \cr
    0 & {1 \over 36} & {1 \over 36} & 0 & {1 \over 9} & 0 & 0 & 0 \cr
    0 & 0 & 0 & {1 \over 36} & 0 & {1 \over 9} & {1 \over 36} & 0 \cr
    0 & 0 & 0 & {1 \over 36} & 0 & {1 \over 36} & {1 \over 9} & 0 \cr
    0 & 0 & 0 & 0 & 0 & 0 & 0 & {1 \over 6} \cr }.
\end{equation}
(We observe, somewhat relatedly, that \.Zyczkowski {\it et al} ``could not
resist the temptation  to investigate the mean value of [the degree of
entanglement] over random density matrices generated'' according to
their measure \cite[App. B]{zycz}.)
\subsection{Monotone metrics} \label{mono}
\begin{figure}
\centerline{\psfig{figure=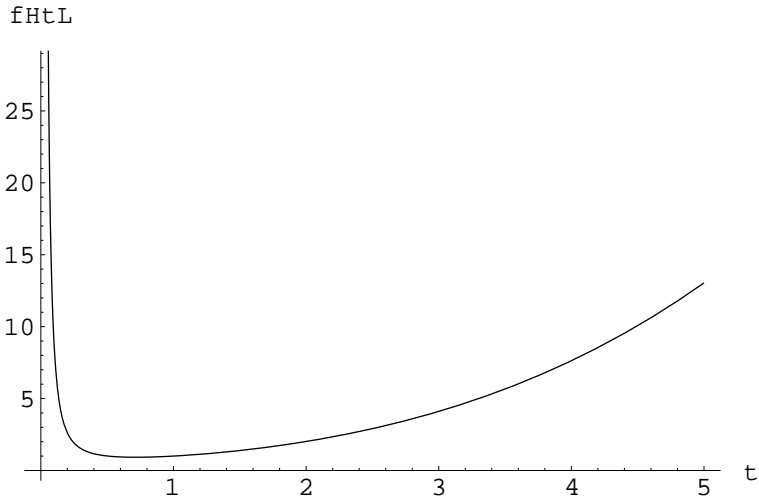}}
\caption{Nonmonotone nature of the indicator
function (\ref{nonmonotone}), corresponding to the probability
distribution (\ref{prob}) with $u=-2$}
\label{fignonmonotone}
\end{figure}

So, we are able to conclude that for the case $N=2$, the measure employed
by \.Zyczkowski {\it et al} does not correspond to that for either the
minimal or monotone metric.
In fact, based on the evidence so far presented, one might conjecture
 that it does not correspond to any monotone metric at all. We say this
based on the proposition that for
any monotone metric one can associate a ``Morozova-Chentsov function'',
expressible as $c(x,y) = 1/y f(x/y)$,
 where $f(1)=1$ and $f(t) =t f(t^{-1})$. For the family of probability
 distributions (\ref{prob}), we have that \cite[eq. (3.17)]{petzsudar}
\begin{equation} \label{monotone}
f(t) = {(1+t)^{2 -2 u} \over 2^{2 -2 u} t^{{1 \over 2} -u}}.
\end{equation}
In the minimal ($u = {1 \over 2}$) and maximal ($u = {3 \over 2}$) cases,
$f(t) = { 1 +t \over 2}$ and 
${2 t \over {1 +t}}$, respectively. Now, both of these functions
are {\it operator monotone} --- which constitutes a necessary
and sufficient condition for the monotonicity of the
associated metric. (A function $f: {\Bbb{R}}^{+}
\to {\Bbb{R}}$ is called operator monotone if the relation
$0 \leq K \leq H$ implies $0 \leq f(K) \leq f(H)$ for any
matrices $K$ and $H$ of any order. Ordinary monotonicity is obtained if
$K$ and $H$ are simply scalar quantities.) But for the choice $u = -2$, 
which yields (\ref{resultm2}) and (\ref{resultm3}), we have that
\begin{equation} \label{nonmonotone}
f(t) = { (1+t)^6 \over 64 t^{{5 \over 2}}},
\end{equation}
 which is not monotone
(Fig.~\ref{fignonmonotone}) nor, {\it a fortiori}, operator monotone.
(``All stochastically monotone Riemannian metrics are characterized by
means of operator monotone functions and \ldots there exists a maximal and
minimal among them'' \cite{petzsudar}.)
\subsection{The subcases $m=4,5,6$}
Continuing on to the scenario $m =4$, 
however, the legitimacy of this particular argument (sec.~\ref{mono}), 
concerning monotonicity, is
undermined by the fact that the $16 \times 16$
 analog of (\ref{resultm2}) and
(\ref{resultm3}) is {\it not}  the same as that achieved using the
probability distribution (\ref{prob}) with $u = - 2$ (or any other specific
value of $u$).
Nevertheless,  these two $16 \times 16$ density matrices share the
same zero-nonzero pattern and possess  two similar sets of three distinct
eigenvalues
(${7 \over 66}, {1 \over 22}, {1 \over 33}$  for the mean matrix
based on (\ref{prob}) setting $u=-2$, and 
${8 \over 75},{2 \over 45},{1 \over 30}$ for the one relying upon the
measure of \.Zyczkowski {\it et al}) of
multiplicities five, nine and two, respectively, corresponding to 
{\it identical}
eigenspaces. (C. Krattenthaler has a formal demonstration
 that averaging over the Bloch sphere with respect to
any {\it spherically-symmetric} probability distribution yields the same
collection of eigenspaces for any $m$.) 

For the cases $m=5$ and 6, we have found results of a similar nature.
Using the $U(2)$
 measure of \.Zyczkowski {\it et al}, the eigenvalues for $m=5$ are
$ {13 \over 180} $, ${1 \over 40}$ and
 $ {1 \over 60} $ (with multiplicities six,
sixteen and ten, respectively) and for $m=6$, they are $ {151 \over 2940}$,
 ${31 \over 2100}$, ${11 \over 1260}$ and ${1 \over 140}$ (with multiplicities
seven, twenty-five, twenty-seven and five, respectively).
(These multiplicities are in conformity with the formula (\ref{multiplicity})
below,
reported in \cite{kratt}.)
The corresponding averaged matrices and those based on (\ref{prob}) share the 
same zero-nonzero patterns, but do not fully match for any particular value
of $u$. (We have not, however, explicitly confirmed that the eigenspaces
are identical, as we anticipate.)

It would be quite interesting, it would appear, for its possible
utility in universal quantum coding \cite{kratt,jozsa}, to obtain a formula for
general $m$
for the entries and eigenvalues of the $2^m \times 2^m$ averaged density
matrix based on the measure of \.Zyczkowski {\it et al}.
(Presumably, the eigenvectors are, for all $m$, those already given by
Krattenthaler and Slater \cite{kratt}.) Then, one could
try to establish the asymptotics of the 
relative entropy with respect to this averaged matrix of the $m$-fold
tensor products of the $2 \times 2$ density matrices.
\subsection{Explicit spin states} \label{explicitspin}

For general $m$, in the framework of \cite{kratt} based on (\ref{prob}),
the subspace spanned by all those eigenvectors associated with the same
($d$-th)
eigenvalue \cite[eq. (2.12)]{kratt},
\begin{equation} \label{eigen}
\lambda_{m,d} = {1 \over 2^{m}} {\Gamma({5 \over 2} -u) \Gamma(2 +m -d -u)
\Gamma(1+d-u) \over \Gamma({5 \over 2} +{m \over 2}-u) \Gamma(1-u)},
\qquad d =0,1,\ldots,\lfloor {m \over 2} \rfloor
\end{equation}
corresponds to those {\it explicit spin states}
 \cite[sec. 7.5.j]{biedenharn} \cite{pauncz1,pauncz2,slatqst}
 with $d$
spins ``up''  or
``down'' (and the other $m-d$ spins, of course, the reverse).
 (Eigenvectors --- linearly independent, but not orthonormalized --- are
enumerated by ``ballot paths'' and given by formula
(2.14) of \cite{kratt}.)	
The $2^{m}$-dimensional Hilbert space can be decomposed into the direct sum
of carrier spaces of irreducible representations of $SU(2) \times S_{m}$.
The multiplicities of the eigenvalues \cite[eq. (2.13)]{kratt},
\begin{equation} \label{multiplicity}
   M_{m,d} = {(m-2 d +1)^2 \over (m+1)} {m + 1 \choose d},\qquad d =0,1,\ldots,
\lfloor {m \over 2} \rfloor
\end{equation}
are the dimensions of the corresponding irreps.
Thus, for $m=4,u=-2$, using (\ref{eigen}) and
(\ref{multiplicity}), we have the results previously mentioned,
$\lambda_{4,0} = {7 \over 66}, \lambda_{4,1}={1 \over 22},
\lambda_{4,2}={1 \over 33}$, and $M_{4,0}=5,
M_{4,1}=9,M_{4,2} =2$.
Employing these two formulas again, we can easily ascertain that
the eigenvalues of the $4 \times 4$ matrix
 (\ref{resultm2}) are ${5 \over 18}$ (threefold) and
${1 \over 6}$ (unrepeated). For the $8 \times 8$ matrix
(\ref{resultm3}), we have the fourfold eigenvalue ${1 \over 6}$
associated with the orthonormalized eigenvectors,
\begin{equation} \label{moreeigen1}
(0,0,0,0,0,0,0,1),\qquad (0,0,0,{1 \over \sqrt{3}},0,{1 \over \sqrt{3}},
{1 \over \sqrt{3}},0),\qquad (0,{1 \over \sqrt{3}},{1 \over \sqrt{3}},0,
{1 \over \sqrt{3}},0,0,0),\qquad (1,0,0,0,0,0,0,0)
\end{equation}
 and the fourfold eigenvalue ${1 \over 12}$, corresponding to
\begin{equation} \label{moreeigen2}
(0,0,0,-{1 \over \sqrt{2}},0,0,{1 \over \sqrt{2}},0), \qquad
(0,0,0,-{1 \over \sqrt{6}},0, {\sqrt{2} \over \sqrt{3}},
-{1 \over \sqrt{6}},0),
\end{equation}
\begin{displaymath}
(0,-{1 \over \sqrt{2}},0,0,{1 \over \sqrt{2}},0,0,0), \qquad
(0,-{1 \over \sqrt{6}},{\sqrt{2} \over \sqrt{3}},0,-{1 \over \sqrt{6}},0,0,0).
\end{displaymath}

Our investigation will now move on to cases for which $N > 2$.
However, in this regard, let us point out that in their discussion of
explicit spin-j states, Biedenharn and Louck
\cite[p. 423]{biedenharn} have
remarked that ``Unfortunately, for 
arbitrary $n$ and $k$, the construction of the basis vectors \ldots has
never been given (fully explicitly), principally because of unsolved problems
relating to the additional labels ($\alpha$), which are required to specify
a basis of the space ${\mathcal{V}}^{(n)}_{k}$,
 and which imply a multiplicity of
occurrence of the irrep \ldots of $SU(2) \times S_{n}$ in the representation.
However, {\it for the case} $k = {1 \over 2}$, it may be proved, using
standard character formulas \ldots, that {\it the representation  \ldots
contains no multiply occurring irreps of $SU(2) \times S_{n}$ \ldots}
For the case of the coupling of $n$ spin-${1 \over 2}$ angular momenta,
the indices ($\alpha$) are not required in the notation \ldots for
the basis vectors.''
(``The so-called missing label problem is a recurring one in the theory of
group and Lie algebra representation theory. To be specific, suppose that
$\lambda$ labels an irrep of a group $G$ and $\nu$ indexes a basis for an
irrep $\kappa$ of a subgroup $H \subset G$. If the $\kappa$ irreps of $H$ 
that occur in the space of the $\lambda$ irreps of $G$ are multiplicity free,
then a basis $ \lbrace | \lambda \kappa \nu \rangle \rbrace$ is uniquely
defined by the $(\lambda \kappa \nu)$ labels. However, if some $\kappa$ irrep 
has a multiple occurrence, an additional multiplicity index $\alpha$ is 
needed to label a complete basis $\lbrace | \lambda \alpha \kappa \nu
\rangle \rbrace$. This is the generic situation'' \cite{rowe}
(cf. \cite{alisauskas}).) 

In a related context, Katriel, Paldus and Pauncz \cite{katriel} in an
article entitled, `` Generalized Dirac Identities and Explicit Relations
between the Permutational Symmetry and the Spin Operators for Systems
of Identical Particles,''
 wrote that,
``the most interesting problem to be considered is associated with the
breakdown of the one-to-one correspondence between the total spin and the
irreducible representations of the symmetric group for systems of identical
particles with an elementary spin $\sigma > {1 \over 2}$ \ldots For two
particles the symmetric group is $S_{2}$ with only two irreducible 
representations, $\mathbf{[2]}$ and $\mathbf{[1^{2}]}$.
 For $\sigma = {1 \over 2}$ these two representations
fully characterize the two possible total spin states $S=0$ and $S=1$,
respectively. However, for two $\sigma = 1$ particles three different total
spin states are possible, two of which are symmetric
 ($\mathbf{[2]}$) and one of which
is antisymmetric ($\mathbf{[1^{2}]}$).
 This is the first example where we no longer
have a one-to-one correspondence between the total spin states and the
irreducible representations of the symmetric group. For three $\sigma =1$
particles the total spin is already insufficient to specify the irreducible
representation; there are two different $S=1$ states, each belonging to a
different irreducible representation. This situation is repeated, more
frequently, for states of four and five particles, but it is only for
six $\sigma =1$ particles that an even more severe labeling problem appears:
namely, two different sets of functions with the same total spin ($S=2$)
correspond to the same irreducible representation, $\mathbf{[4,2]}$.
 For $\sigma
= {3 \over 2}$ the same labeling difficulties are encountered. Already
for four such particles, more than one state with a given total spin
corresponds to the same irreducible representation. For $\sigma =2$ this
degeneracy is present for three particles already. In other words, the
representation of the symmetric group generated by the spin eigenfunctions
is a reducible one, and the irreducible components can have multiplicities
larger than 1'' (cf. \cite{kent87,kent89,kent93}).

The abstract to this  paper of Katriel, Paldus and Pauncz \cite{katriel}
reads: ``The well-known one-to-one correspondence between the eigenstates
of the total spin for a system of spin-${1 \over 2}$ particles and 
irreducible representations of the symmetric group with up to two rows in
the Young shape is the basis of interesting formal developments in quantum
chemistry and in the theory of magnetism. As an explicit manifestation of
this correspondence the class operators of the symmetric group are
demonstrated to be expressible in terms of the total spin operator.
This correspondence does not hold for higher elementary spins. The extension
to arbitrary spin is investigated using Schr\"odinger's generalization
of the Dirac identity, which expresses the transposition in terms of 
two-particle spin operators. It is shown  that additional operators,
which for $\sigma = {1 \over 2}$ reduce to the total spin operator,
are needed for a complete classification. Some aspects of the formalism
are developed in detail for $\sigma =1$. In this case a classification
identical with that provided by the irreducible representations of the
symmetric group is obtained in terms of the eigenstates of two commuting
operators, one of which is the total spin operator.''

It appears that the research reported below is somewhat similar in nature
to that of Katriel, Paldus, and Pauncz, but with the focus
now on the eigenstates not of the $N \times N$
 operators $O_{k}$ (in their notation),
where \cite[eq. (38)]{katriel} ($s_{i} = (s_{x i},s_{y i},s_{z i})$ being
a one-particle spin operator)
\begin{equation}
O_{k} = O_{k}(N) = \sum_{i < j}^{N} (s_{i} s_{j})^k,
\end{equation}
  but on the eigenstates
of the $N^{m} \times N^{m}$ mean density matrix  obtained by averaging with
respect to the (normalized) $U(N)$-invariant measure.
(``We have not been able to work out the general commutation relation
between $O_{k}$ and $O_{l}$ for arbitrary $k$ and $l$ larger than 1, but
we would like to conjecture that {$O_{n};n=1,2,\ldots,2 \sigma$} form a
set of commuting operators for arbitrary $\sigma$. The missing proof of this
conjecture is the most obvious loose end in the present investigation''
\cite{katriel}.)
\section{THE CASE $N=3$} \label{N3case}
\subsection{The subcase $m=1$}
We have performed a series of
 computations in the framework of \.Zyczkowski
{\it et al} \cite{zycz} for the case $N =3$,
 utilizing the {\it Euler-angle} parameterization
of $SU(3)$, along with the invariant volume element,
 recently given by Byrd \cite[eq. (1) and p. 14]{byrd}
(cf. \cite[eq. (17)]{byrd2}
and \cite[p. 5]{byrd3}). Averaging the $3 \times 3$ density matrices
accordingly,
we obtained the simple intuitive result (cf. (\ref{mixed1})),
\begin{equation} \label{mixed2}
\pmatrix{{1 \over 3} & 0 & 0 \cr
          0 & {1 \over 3} & 0 \cr
          0 & 0 & {1 \over 3} \cr}.
\end{equation}
We have not, for the $3 \times 3$ density matrices
 been able to perform a strictly comparable averaging, utilizing the
normalized volume elements of the maximal and monotone metrics, but certain
interesting
 results based on them have, nevertheless,
 been obtained \cite{slat2,slat5}.
(We first
 note, analogously to the $2 \times 2$ density matrices, that the integral 
over the {\it entire} eight-dimensional convex set of
the volume element of the maximal monotone metric {\it diverges}.
For the minimal monotone metric, on the other hand,
 such divergence appears not to
occur. However, it is more problematical in nature, in that
 it appears not possible to symbolically integrate
 the volume element completely over the eight dimensions \cite{slat5}.)  
In particular, in \cite{slat5}, using a {\it double}-limiting argument
(necessitated by the non-normalizability of the volume element of
the maximal monotone metric),
a {\it six}-dimensional {\it marginal} probability 
distribution was found based on  a set
of eight (separable)
 variables parameterizing the $3 \times 3$ density matrices.
Its {\it two}-dimensional marginal probability over the simplex 
spanned by the diagonal entries
($a,b,c$) was
\begin{equation} \label{marginalprob}
{15 (1-a) \sqrt{a} \over 4 \pi \sqrt{b} \sqrt{c}}.
\end{equation}
The original symmetry between these three variables was {\it broken} by
a series of transformations --- suggested by work of Bloore
 \cite{bloore} --- employed to obtain the eight variables,
the separation of which was required in order to perform the necessary
integrations to obtain (\ref{marginalprob}).
It is, then, natural to associate the variable $a$ with the {\it middle}
of the three levels (the one inaccessible to a spin-1 photon, due to its
masslessness).
The expected values with respect to 
(\ref{marginalprob}) are
$\langle a \rangle = {3 \over 7},\langle b \rangle =\langle c \rangle
={2 \over 7}$. We, of course, note that neither of these values equals
${1 \over 3}$ as in (\ref{mixed2}), based on the approach of
\.Zyczkowski {\it et al} \cite{zycz}.
\subsection{The subcase $m=2$} \label{N3m2}

We have also conducted the same form of averaging as used to get
(\ref{mixed2}) for the
$9 \times 9$ density matrices (the ``two-trit'' case
\cite{horod5,horod6,linden}) obtained by taking the twofold tensor
products ($m=2$) of the $3 \times 3$ density matrices 
($N=3$). The result obtained is
the symmetric matrix (the $U(3)$ analog of the
$4 \times 4$ matrix   (\ref{resultm2}), that was based on  $U(2)$
 (cf. \cite[p. 423]{biedenharn}))
\begin{equation} \label{ninebynine}
\pmatrix{{1 \over 8} & 0 & g & 0 & 0 & {10 g \over 3} & g  & {10 g \over 3}
           & 0\cr
         0 & {5 \over 48} & -2 g  & {1 \over 48} & 0 & -2 g & -2 g & -2 g
          & 0 \cr
         g & -2 g & {5 \over 48} & -2 g & {10 g \over 3} & 0 & {1 \over 48} & 0 & g  \cr
         0 & {1 \over 48} & -2 g & {5 \over 48} & 0 & -2 g & -2 g & -2 g
          & 0 \cr
         0 & 0 & {10 g \over 3} & 0 & {1 \over 8} & g & {10 g \over 3} & g
 & 0 \cr
         {10 g \over 3} & -2 g & 0 & -2 g & g  & {5 \over 48} & 0 &
 {1 \over 48} & g \cr
         g & -2 g & {1 \over 48} & -2 g & {10 g \over 3}
 & 0 & {5 \over 48} & 0 & g \cr
         {10 g \over 3} & -2 g & 0 & -2 g & g & {1 \over 48} & 0 & {5 \over 48} & g \cr
         0 & 0 & g & 0 & 0 & g & g & g & {1 \over 8} \cr},
\end{equation}
where $g = {1 \over 864 \pi} \approx .000368414$. (The off-diagonal entries,
${1 \over 48}$, lacking the constant $\pi$, correspond to the expected
value of the product of symmetrically located off-diagonal
 entries --- $a_{ij} a_{ji}$ --- in
the underlying $3 \times 3$ density matrix [$a_{ij}$].)
Tracing the $9 \times 9$ density matrix (\ref{ninebynine}) over one of the two
constituent subsystems, we obtain the $3 \times 3$ 
diagonal density matrix 
(\ref{mixed2}) with entries equal to ${1 \over 3}$.

Of the nine eigenvalues of (\ref{ninebynine}), three  are ${1 \over 12}$,
two  are ${1 \over 8}$,
and the other four (unrepeated) ones can be paired as
\begin{equation} \label{unrepeated}
\lambda_{6} = {1 \over 8} + {7 \over 1296 \pi \sqrt{2}}
 \approx .126216, \qquad
\lambda_{7} = {1 \over 8} - {7 \over 1296 \pi \sqrt{2}} \approx .123784,
\end{equation}
and
\begin{displaymath}
\lambda_{8} = {1 \over 8} + {\sqrt{331} \over 1296 \pi \sqrt{2}} \approx
.12816, \qquad
\lambda_{9} = {1 \over 8} - {\sqrt{331} \over 1296 \pi \sqrt{2}} \approx
.12184.
\end{displaymath}
The three-dimensional subspace is spanned by the orthonormal set of 
eigenvectors
\begin{equation} \label{eigentriple}
(0,0,0,0,0,-{1 \over \sqrt{2}},0,{1 \over \sqrt{2}},0),
 \qquad (0,0,-{1 \over \sqrt{2}},0,0,0,{1 \over \sqrt{2}},0,0), \qquad
 (0,-{1 \over \sqrt{2}},0,{1 \over \sqrt{2}},0,0,0,0,0),
\end{equation}
while the two-dimensional subspace is generated  by
\begin{equation} \label{eigendouble}
(0,{1 \over 3 \sqrt{2}},0,{1 \over 3 \sqrt{2}},0,0,0,0,{2  \sqrt{2} \over 3}),
\qquad ({9 \over \sqrt{331}},{26 \over 3
 \sqrt{331}},0,{26 \over 3 \sqrt{331}},
{9 \over \sqrt{331}},0,0,0,-{13 \over 3 \sqrt{331}}).
\end{equation}
The remaining four one-dimensional eigenspaces 
(corresponding to the sequence of eigenvalues (\ref{unrepeated}),
which can be viewed as providing ``labels'' to the
eigenspaces) are given  by
\begin{equation} \label{eigenisolated}
(-{1 \over 2},0,-{1 \over 2 \sqrt{2}},0,{1
 \over 2},{1 \over 2 \sqrt{2}},-{1 \over 2 \sqrt{2}},{1 \over 2 \sqrt{2}},0),
 \qquad
({1 \over 2},0,-{1 \over 2 \sqrt{2}},0,-{1 \over 2},
{1 \over 2 \sqrt{2}},-{1 \over 2 \sqrt{2}},{1 \over 2 \sqrt{2}},0),
\end{equation}
\begin{displaymath}
({13 \over 2 \sqrt{331}},-{6 \over \sqrt{331}},-{1 \over 2 \sqrt{2}},-{6 \over
 \sqrt{331}},{13 \over 2 \sqrt{331}},-{1 \over 2 \sqrt{2}},-{1 \over 2
 \sqrt{2}},
-{1 \over 2 \sqrt{2}},{3 \over \sqrt{331}}),
\end{displaymath}
\begin{displaymath}
({13 \over 2 \sqrt{331}}, -{6 \over \sqrt{331}},{1 \over 2 \sqrt{2}},
-{6 \over \sqrt{331}},{13 \over 2 \sqrt{331}},{1 \over 2 \sqrt{2}},
{1 \over 2 \sqrt{2}},{1 \over 2 \sqrt{2}},{3 \over \sqrt{331}}).
\end{displaymath}
We note that the constant $\pi$, appearing 
throughout the
$9\times 9$ averaged matrix
(\ref{ninebynine}) and in the four eigenvalues
 (\ref{unrepeated}),  is absent --- similarly to the parameter $u$
in the $2^{m} \times 2^{m}$ analyses of Krattenthaler
and Slater \cite{kratt} --- from the set of nine orthonormalized eigenvectors
(\ref{eigentriple}), (\ref{eigendouble}) and (\ref{eigenisolated}).
Thus, we are able, by replacing $\pi$ 
in (\ref{ninebynine})
 by a free parameter ($v$), to obtain a one-parameter
 family of $9 \times 9$
matrices, all the members of which have the same sets of
 eigenvectors (\ref{eigentriple}),
(\ref{eigendouble}) and (\ref{eigenisolated}).
(We might speculate that there exists a family of probability distributions,
parameterized by $v$, which would give these $9 \times 9$ density matrices,
as their expected values, as we now know is, in fact, the case for 
the particular value $v=\pi$ [ cf. sec.~\ref{secparam}].)
For $|v| \geq {\sqrt{331} \over 162 \sqrt{2}} \approx .0794116$, the
$9 \times 9$ matrix has nonnegative eigenvalues,
 so it is, then, a {\it density}
matrix.
If one sets  $v$ to either ${\sqrt{331} \over 54 \sqrt{2}} \approx .238235$
or ${7 \over 54 \sqrt{2}} \approx .091662$, two of the unrepeated
eigenvalues become ${1 \over 12}$ and ${1 \over 6}$. As $v$ tends to 
$\pm \infty$, the four unrepeated eigenvalues 
(\ref{unrepeated}) all approach
${1 \over 8}$. So, in that limit, we obtain a sextet and an ``antitriplet''
\cite[p. 311]{greiner}. The antitriplet is spanned by the vectors
(\ref{eigentriple}), while (replacing (\ref{eigendouble}) and
(\ref{eigenisolated})), the sextet is spanned by
\begin{equation} \label{eigensextet}
(0,0,0,0,0,0,0,0,1),\qquad
 (0,0,0,0,0 {1 \over \sqrt{2}},0,{1 \over \sqrt{2}},0),\qquad
(0,0,{1 \over \sqrt{2}},0,0,0,{1 \over \sqrt{2}},0,0),
\end{equation}
\begin{displaymath}
(0,0,0,0,1,0,0,0,0),\qquad
 (0,{1 \over \sqrt{2}},0,{1 \over \sqrt{2}},0,0,0,0,0), \qquad
(1,0,0,0,0,0,0,0,0).
\end{displaymath}
``We know that the product of two $SU(3)$ triplets decomposes into a sextet
and an antitriplet. The sextet is symmetric, whereas the antitriplet is
antisymmetric. Therefore, both representations can be contained with the
baryon multiplet which has mixed symmetry, but only the sextet can be
contained in the multiplet of baryon resonances which corresponds to a totally
symmetric representation'' \cite[p. 388]{greiner}.
\subsection{The subcase $m=3$}
For the subcase $m=3$ of $N=3$,  the diagonal
entries of the $27 \times 27$ density matrix, obtained by averaging
 according to the measure of
\.Zyczkowski {\it et al} \cite{zycz}, are
\begin{equation}
(\alpha,\beta,\beta,\beta,\beta,\gamma,\beta,\gamma,\beta,\beta,\beta,
\gamma,\beta,\alpha,\beta,\gamma,\beta,\beta,
\beta,\gamma,\beta,\gamma,\beta,\beta,\beta,\beta,\alpha),
\end{equation}
where $\alpha ={ 31 \over 600}
\approx .0516667,\beta={11 \over 300} \approx .0366667,
\gamma={37 \over 1200} \approx .030833$.
We have been able to determine  the {\it off}-diagonal entries
(of which, 272 are zero),
as well. For
 instance, the (1,3) entry along with twenty-three others have the
value ${7  \over 8640 \pi}$. Of the non-zero off-diagonal entries, 
most were of such a form, that is ${ k \over l \pi}$. However, there were
also
 three distinct {\it rational} numbers: 
${3 \over 400}$ (thirty-six occurrences),${7 \over 1200}$
(eighteen occurrences) and ${1 \over 600}$ (twelve occurrences).
If we denote the entries of the underlying $3 \times 3$ density matrix
by $a_{ij}$ ($i,j=1,2,3$), then the cells  in the $27 \times 27$
density matrix yielding ${3 \over 400}$ corresponded to entries of the
form $a_{ii} a_{ij} a_{ji}$, those yielding ${7 \over 1200}$
corresponded to entries of the form $ a_{ii} a_{jk} a_{kj}$, and those
yielding ${1 \over 600}$ were associated
with cells of the type $a_{ij} a_{jk} a_{ki}$
(where $i,j,k$ now refer to {\it distinct} values of 1, 2 or 3).

Seven of the twenty-seven eigenvalues were expressible as rational
numbers. These were
\begin{equation}
{1 \over 60} \approx .0166667 \quad (\mbox{isolated}), \qquad
{7 \over 240} \approx
.0291667  \quad (\mbox{fourfold}),\qquad 
{31 \over 600}  
 \approx .0516667 \quad (\mbox{twofold}).
\end{equation}
Among the remaining eigenvalues, there were three pairs, corresponding to the
roots ($\lambda$) of the
sixth-degree even polynomial, where $y = (7 -240 \lambda) \pi$,
\begin{equation} \label{EQ1}
5504 - 317043 y^{2}  + 4986360 
y^{4}  -19131876 y^{6} =0.
\end{equation}
The eigenvalues ($\lambda$) are {\it explicitly} expressible as
\begin{equation} \label{pairedeigenvalues}
{7 \over 240} - {1 \over 720 \pi \sqrt{2}} \approx .0288541,\qquad
{7 \over 240} + {1 \over 720 \pi \sqrt{2}} \approx .0294793,
\end{equation}
\begin{displaymath}
{7 \over 240} - {1 \over 405 \pi \sqrt{2}} \approx .0286109,\qquad
{7 \over 240} + {1 \over 405 \pi \sqrt{2}} \approx .0297224,
\end{displaymath}
and
\begin{displaymath}
{7 \over 240} - {\sqrt{43} \over 6480 \pi \sqrt{2}} \approx .0289389, \qquad
{7 \over 240} + {\sqrt{43} \over 6480 \pi \sqrt{2}} \approx .0293994.
\end{displaymath}
There were also eight isolated eigenvalues ($\lambda  = $ .0495426,
 .0496514, .0500807, .0515902, .0517431,
.0532526, .0536819, .0537907), corresponding to the roots of the eighth-degree
even polynomial, where $y = (31 -600 \lambda) \pi$,
\begin{equation} \label{EQ2}
383127080633857021  - 18544605647405907654  y^{2}+
4501753947892101744 y^{4}
\end{equation}
\begin{displaymath}
 - 351843587054438400 
y^{6} + 8926168066560000 y^{8} = 0.
\end{displaymath}
The smallest and largest  of these eight isolated eigenvalues are
\begin{equation} \label{unpairedeigenvalues}
{31 \over 600} - { \sqrt{2337181} \over 162000 \pi \sqrt{2}} \approx
.0495426, \qquad
{31 \over 600} + {\sqrt{2337181} \over 162000 \pi \sqrt{2}} \approx
.0537907.
\end{equation}
If we replace the constant $\pi$ by the variable $v$
(as was done in the $N=3, m =2$ case of sec.~\ref{N3m2}
 [cf. (\ref{eigensextet})]),
 then the seven 
rational eigenvalues are unchanged and the other twenty
 can be obtained from the
solutions ($\lambda$) of (\ref{EQ1}) and (\ref{EQ2})
 using now $y = (7-240 \lambda) v$
and $y= (31 -600 \lambda) v$.
In the limit $v \rightarrow \pm \infty$, the eigenvalues of the matrix
degenerate to
${1 \over 60}$ (isolated), ${7 \over 240}$ (sixteenfold) and
${31 \over 600}$ (tenfold), in agreement with the cardinalities of the
multiplets of $SU(3)$ constructed from three particles
 \cite[exer. 9.5]{greiner},\cite[eq. (11.5.4)]{leader}.

The eigenvector corresponding to the lowest-lying (isolated) eigenvalue
(${1 \over 60}$) is 
\begin{equation} \label{iso1}
(0,0,0,0,0,-a,0,a,0,0,0,a,0,0,0,-a,0,0,0,-a,0,a,0,0,0,0,0),
\end{equation}
where $a = {1 \over \sqrt{6}}$.

\subsection{The subcase $m=4$}
We have ascertained that on the diagonal of the
$81 \times 81$ density matrix averaged according to the measure of
\.Zyczkowski {\it et al} \cite{zycz},
 there are three occurrences of ${7 \over 300} \approx
.0233333$, eighteen of ${11 \over 900} \approx .0122222$, twenty-four of
${17 \over 1200} \approx .0141667$ and thirty-six of
${37 \over 3600} \approx .0102778$. We have only so far, however, been able
to determine a relatively
 small number of the off-diagonal entries. For example,
the (1,3)-entry has the value ${41\over 86400 \pi}$, the
(2,3)-entry, ${-{43 \over 54000 \pi}}$, and the (11,12)-entry,
${-{37 \over 108000 \pi}}$.

\section{THE CASE $N= 2 \times 3$} \label{N6case}
\subsection{The subcase $m=1$}
Let us consider the tensor product of an arbitrary 
 $2 \times 2$ density matrix and an arbitrary 
$3 \times 3$ density matrix (cf. \cite{horo8}).
 Then, we average the $6 \times 6$
result (corresponding to the subcase $m=1$) over the convex sets of
two-dimensional and three-dimensional density matrices, using the
$U(2) \times U(3)$ product
 measure, in the natural extension of the work of \.Zyczkowski {\it et al}
\cite{zycz}. We obtain a diagonal matrix with entries ${1 \over 6}$,
 which is simply the same as the
 tensor product of (\ref{mixed1}) and (\ref{mixed2}),
evidence of the statistical independence of the constituent density
matrices.
\subsection{The subcase $m=2$}
Now, however, let us similarly average the twofold tensor products 
($m=2$) of
these $6 \times 6$ density matrices with themselves.
Then, the eigenvalues of the averaged $36 \times 36$ density matrix 
(966 of the 1296 entries being zero) are
${1 \over 72} \approx .0138889$ (multiplicity three), ${1 \over 48}
\approx .020833$ (two),
${5 \over 216} \approx .0231481$ (nine), ${5 \over 144}
\approx .0347222$ (six) and the remaining ones
(all explicitly
 expressible as fractions involving square roots and $\pi$) are .0203067
(one), .0206307
(one), .021036 (one), .0213599 (one), .0338445 (three), .0343845
(three), .0350599 (three) and .0355999 (three).
For instance, the last of these values is an approximation 
(cf. (\ref{unrepeated})) to
${5 \over 144} + {5 \sqrt{331} \over 23328 \pi \sqrt{2}}$.
If, as previously, we replace the occurrences of $\pi$ in 
(the denominators of certain entries of) the averaged matrix
by a parameter $v$ and let $v \rightarrow \infty$ (or, equivalently,
simply set the entries in question to zero), we obtain the eigenvalues
${1 \over 72}$ (multiplicity three), ${1 \over 48}$ (six), ${5 \over 216}$
(nine) and ${5 \over 144}$ (eighteen). 
For the first of the corresponding four eigenspaces, all the nonzero entries of
the (three) orthonormalized 
spanning eigenvectors were $\pm {1 \over 2}$, for the
second and third eigenspaces,
 $\pm {1 \over 2}$ or $\pm {1 \over \sqrt{2}}$, and for
the (dominant) fourth, the entries were
 1 or ${1 \over 2}$ or ${1 \over \sqrt{2}}$.
\section{THE CASES $N=2 \times 3 \times 2$ and $N=
3 \times 2 \times 2$} \label{N144one}
Let us continue further along the lines of the immediately preceding
analysis (sec.~\ref{N6case}). We construct a $12 \times 12$ density matrix
by taking the ordered tensor product of arbitrary $2 \times 2$,
$3 \times 3$ and $2 \times 2$ ones. Then, by taking the tensor product of
the result with itself, we obtain
a $144 \times 144$ density matrix, which we average with respect to
the product measure for $U(2) \times U(3) \times U(2)$.
The mean density matrix obtained had 16,886 of its 20,736 entries, zero.
Let us, first, report the structure of the eigenvalues of this matrix, if
we replace the occurrences in its off-diagonal cells of $\pi$ by
 a parameter $v$, which we then let
go to infinity, as we have previously in this series of analyses. There are,
then, six eigenspaces, corresponding to the eigenvalues ${1 \over 360}$
(mulitplicity six), ${1 \over 270}$ (twenty-seven), ${1 \over 240}$
(twelve),${1 \over 180}$ (fifty-four), ${2 \over 225}$ (fifteen),
${1 \over 75}$ (thirty).

Leaving the constant $\pi$ unaltered, we obtain 
eighteen, rather than six, eigenspaces. The multiplicities
of the eigenvalues ${1 \over 360},{1 \over 270}$ and
 ${2 \over 225}$ are the same as
in the $v \rightarrow \infty$ analysis, while that of ${1 \over 240}$ is
reduced from twelve
 to four, ${1 \over 180}$ from fifty-four
 to eighteen and ${1 \over 75}$ from thirty to ten.
The manner in which these reductions in
dimensionality take place can be seen by
examining the orthogonal
 64-dimensional Hilbert space. It is composed of four eigenspaces
of dimension nine (corresponding to the eigenvalues
 ${1 \over 180} \pm  {7 \over
29160 \pi \sqrt{2}}$ and ${1 \over 180} \pm  {\sqrt{331} \over 29160 \pi
\sqrt{2}}$), four  of dimension five (associated with the 
eigenvalues ${1 \over 75} \pm
{\sqrt{331} \over 12150 \pi \sqrt{2}}$ and
${1 \over 75} \pm {7  \over 12150   \pi \sqrt{2}}$) and four of dimension
two (corresponding to the eigenvalues
 ${1 \over 240} \pm {7  \over 38880  \pi \sqrt{2}}$
and ${1 \over 240} \pm {\sqrt{331} \over 38880 \pi \sqrt{2}}$).
Of course, it is easily noted that $4 \times 9 = 54-18$, $4 \times 5 = 30-10$
and $ 4 \times 2 = 12-4$.

We have also conducted a parallel series of analyses for $N =3 \times 2
\times 2$, thus, ordering the three density matrices in the initial
tensor product differently.
We, of course, again obtain a $144 \times 144$ mean density matrix
(having, once again, 16,886  of its 20,736 entries, equal to zero).
In the limit, $v \rightarrow \infty$, an eigenanalysis yielded precisely
the same results as the corresponding analysis in the $2 \times 3 \times 2$
case.
And in fact, leaving the constant $\pi$ unaltered, we obtained the same
set of eighteen distinct eigenvalues and associated multiplicities as
in that case too.
\section{THE CASE  $N=4$} \label{secN4}
For this analysis, we rely upon the (Hurwitz/Euler-angle) parameterization,
together with the accompanying Haar measure, for $U(N)$,
for general $N$, given by
\.Zyczkowski and Ku\'s \cite[eqs. (3.1) -(3.5)]{kus}. For the subcase
$m=1$, the averaged $4 \times 4$ diagonal matrix has non-zero entries equal
to ${1 \over 4}$. For the subcase $m=2$, the averaged matrix had the
diagonal entries
\begin{equation} \label{diagonal16}
(\alpha,\beta,\gamma,\kappa,\beta,\alpha,\gamma,\kappa,\gamma,\gamma,
\epsilon,\zeta,\kappa,\kappa,\zeta,\eta),
\end{equation}
where $ \alpha = {2327 \over 32400}, \beta = {3947 \over 64800}, 
\gamma = {1759 \over 28800}, \kappa =  {971 \over 17280}, \epsilon
= {1583 \over 21600},\zeta= {2357 \over 43200}$ and $ \eta = {299 \over
3600}$. 
The only non-zero off-diagonal entries (in the {\it symmetric} 
mean $16 \times 16$  density matrix) were: ${ 707 \over 64800} $
in the (2,5)-cell;
${319 \over 28800}$ in the (3,9) and (7,10)-cells; ${107 \over 17280}$ in
the (4,13) and (8,14)-cells; and ${ 197 \over 43200}$ in the (12,15)-cell.
We note the absence of the constant $\pi$, in contrast to the scenarios of
secs.~\ref{N3case}, \ref{N6case} and \ref{N144one}, in which
$3 \times 3$ density matrices were incorporated.

There were seven distinct eigenvalues : ${1 \over 20} = .05$ (having
multiplicity six), ${1277 \over 21600} \approx .0591204$ (isolated),
${539 \over 8640} \approx .0623843$ (two), ${2327 \over 32400} \approx
.071821$ (three), ${1039 \over 14400} \approx .0721528$ (two),
${1583 \over 21600} \approx .073287$ (isolated) and 
${299 \over 3600} \approx .0830556$ (isolated)
(cf. \cite[exer. 11.6, eq. (1)]{greiner}).
(The multiplets of a two-particle systems in the group $SU(4)$ are a
{\it sextet} and a decuplet \cite[ex. 11.8(1)]{greiner}.)
Orthonormal bases for the seven eigenspaces are easily constructed,
in which each eigenvector
 has fourteen or fifteen components zero, and the others
either equal to 1 or $\pm {1 \over \sqrt{2}}$.

\section{THE CASE $N=5$}

For the case, $N=5,m=2$, we have only so far been able to determine that
the first two diagonal entries of the mean $25 \times 25$ density matrix
are ${2 \over 45}$ and ${7 \over 180}$ (which must also be the value of
the sixth diagonal entry).
We did in fact  pursue a parallel (and more
computationally manageable) analysis, in which the $5 times 5$
{\it unitary} matrices were 
replaced by the $5 \times 5$ {\it orthogonal} matrices, and the appropriate
Haar measure\cite[App. A]{kus}, then,  used. However, the
 mean density matrix for the case $m=1$ was
found to be null.
\section{PARAMETERIZED FAMILIES OF $N^{m} \times N^{m}$
 MEAN DENSITY MATRICES} \label{secparam}

In our previous analyses, we have employed the measures introduced by
\.Zyczkowski {\it et al} \cite{zycz} to find, for a specific choice of
the dimension $N$ and power $m$, a particular mean density matrix.
However, it should be noted that the imposition by \.Zyczkowski {\it et al}
of a {\it uniform} measure on the $(N-1)$-dimensional simplex spanned by the
eigenvalues ($e_{1},e_{2},\ldots,e_{N}$) is somewhat arbitrary in nature.
In fact, Bayesian principles  would suggest that one
might employ instead a Dirichlet distribution,
\begin{equation} \label{dirichletdist}
{\Gamma{(N-q_{1} -q_{2} - \ldots - q_{N})} \over \Gamma{(1-q_{1})}
\Gamma{(1-q_{2})} \ldots \Gamma{(1-q_{N})}} e_{1}^{-q_{1}}
e_{2}^{-q_{2}} \ldots e_{N}^{-q_{N}}
\end{equation}
with its $N$ parameters ($q_{1},q_{2},\ldots,q_{N}$) all
 set to ${1 \over 2}$ \cite[eq. (3.7)]{bernardoberger}.
(The uniform distribution is obtained by setting  the parameters all to
zero.)
This observation has led us to expand our analysis to include (symmetric)
Dirichlet distributions with all $N$ parameters set equal to
$q$. (We also note that in  the analysis of Krattenthaler and Slater
\cite{kratt} a one-parameter ($u$) family of probability distributions
(\ref{prob}) was utilized to obtain a one-parameter family of
$2^{m} \times 2^{m}$ mean density matrices.)

\subsection{$N=3,m=2$}
We have been able to implement this approach for the case $N=3,m=2$
of sec.~\ref{N3m2}.
All the members of the
 one-parameter family of $9 \times 9$
 density matrices obtained possess the same
zero-nonzero pattern as (\ref{ninebynine}).
The diagonal entries ${1 \over 8}$ and ${5 \over 48}$ 
(corresponding, as noted, to $q=0$) are replaced, now, by the
more general entries, ${3 -2 q \over 6 (4 -3 q)}$ and 
${5 -4 q \over 12 (4 -3 q)}$, respectively.
The constant $g$ is replaced by ${1 \over 216 \pi (4 -3 q)}$ and
the off-diagonal entry, ${1 \over 48}$ by ${1 \over 12 (4 -3 q)}$.
Instead of the three eigenvalues ${1 \over 12}$, we have
${1 \over 9} - {1 \over 9 (4 -3 q)}$ and in place of the double
eigenvalue ${1 \over 8}$, there is ${1 \over 9} + {1 \over 18 (4 -3 q)}$.
The two pairs of unrepeated eigenvalues (\ref{unrepeated}) take the
more general form,
\begin{equation} \label{unrepeatedgeneral}
\lambda_{6} = {3 -2 q \over 6 (4 -3 q)} + {7 \over 324 
(4 -3 q) \pi \sqrt{2}},\qquad
\lambda_{7} = {3 -2 q \over 6 (4 -3 q)} - {7 \over 324
(4 -3 q) \pi \sqrt{2}},
\end{equation}
\begin{displaymath}
\lambda_{8} = {3 -2 q \over 6 (4 -3 q)} + {\sqrt{331}
\over 324 (4 -3 q) \pi \sqrt{2}},\qquad
\lambda_{9} = {3 -2 q \over 6 (4 -3 q)} - {\sqrt{331}
\over 324 (4 -3 q) \pi \sqrt{2}}.
\end{displaymath}
The corresponding eigenspaces are precisely the same as those given by
(\ref{eigentriple})-(\ref{eigenisolated}).

\subsection{$N=4,m=2$}

We have pursued an analogous strategy for the $N=4,m=2$ analysis of
sec.~\ref{secN4}, setting the four parameters of the Dirichlet distribution
(\ref{dirichletdist}), again, all to $q$. The zero-nonzero pattern was
the same as previously reported (that is, for the 
particular instance, $q=0$).
Then, the diagonal entries --- as denoted in (\ref{diagonal16}) --- took
 the form,
\begin{equation}
\alpha = {2327 - 1620 q \over 6480 (5 - 4 q)},\qquad 
\beta={3947 -3240 q \over 12960 (5 -4 q)},\qquad
\gamma= {1759 - 1440 q \over 5760 (5 -4 q)},\qquad \kappa = 
{971 -864 q \over 3456 (5 - 4 q)},
\end{equation}
\begin{displaymath}
 \epsilon = 
{1583 -1080 q \over 4320 (5 -4 q)},\qquad \zeta= {2357 - 2160 q \over
8640 (5 -4 q)},\qquad \eta = {299 -180 q \over 720 (5 -4 q)}.
\end{displaymath}
The (2,5)-cell is now $ {707 \over 12960 (5 -4 q)}$, while the (3,9)
and (7,10)-entries are ${319 \over 5760 (5 - 4 q)}$. Also, the
(4,13) and (8,14)-cells are ${107 \over 3456 (5 -4 q)}$ and the
(12,15)-entry is ${197 \over 8640 (5 -4 q)}$.

The sixfold eigenvalue ${1 \over 20}$ now takes the more general form,
${1 \over 16} - {1 \over 16 (5-4 q)}$, the isolated
eigenvalue ${1277 \over 21600}$ becomes
${1 \over 16}- {73 \over 4320 (5-4 q)}$, while the twofold eigenvalue 
${539 \over 8640}$ is transformed to ${1 \over 16} -
{1 \over 1728 (5 -4 q)}$.
Additionally, the threefold eigenvalue ${2327 \over 32400}$ becomes
${1 \over 16} + {151 \over 3240 (5 -4 q)}$, the twofold eigenvalue ${1039 \over
14400}$ changes to ${1 \over 16} + {139 \over 2880 (5-4 q)}$,
the isolated eigenvalue ${1583 \over 21600}$ is converted to
${1 \over 16}+ {233 \over 4320 (5 -4 q)}$ and the isolated eigenvalue
${299 \over 3600}$ is transformed to ${1 \over 16} + {37 \over 360 (5 -4 q)}$.
The associated eigenspaces were as reported above.

\subsection{$N=4,m=3$}
We have begun to investigate this particular scenario and have established,
among other items,
that the (52,52)-entry of the $64 \times 64$ mean density matrix is
\begin{equation} \label{eigen5252}
{43581 +10 q (2160 q -6283) \over 172800 (5 -4 q) ( 3 - 2 q)},
\end{equation}
the (52,61)-cell is
\begin{equation}
{5026 -2675 \over 172800 (5 -4 q) (3 -2 q)},
\end{equation}
while the (53,53)-entry is 
\begin{equation} \label{eigen5353}
{58387 + 6 q (6480 q - 15979) \over 311040 (5 -4 q) (3 -2 q)}.
\end{equation}

\subsection{$N=2$}

We have also for the case, $N=2$, been able to obtain a {\it two}-parameter
family of $2^{m} \times 2^{m}$ mean density matrices by replacing the
uniform measure of \.Zyczkowski {\it et al} by an (asymmetric) Dirichlet
(or beta) distribution (\ref{dirichletdist}), with its two parameters
$q_{1}$ and $q_{2}$ not necessarily being equal.
Nevertheless, the decompositions of $2^{m}$-dimensional Hilbert space into 
the associated eigenspaces of the mean density matrices
appear to be  of the very
 same nature as reported in sec.~\ref{explicitspin}.
If we do equate the two parameters
($q_{1}=q_{2} \equiv q$), then for the case, $m=2$, we obtain
the isolated eigenvalue ${ 1- q \over 2 (3 -2 q)}$ and the threefold
eigenvalue ${5 -3 q \over 6 (3 -2 q)}$.
For $m =3$, we have two quadruplets, ${2 - q \over 4 (3 -2 q)}$
and ${ 1- q \over 4 (3 -2 q)}$.

Analogously to what was done in \cite{slatbures}
for the $2^{m} \times 2^{m}$ density
matrices, one might investigate the nature
of the Bures distance between two mean density matrices 
(for $N > 2$) corresponding to
nonidentical
 sets of values of the $N$ parameter(s) of the Dirichlet distribution
(\ref{dirichletdist}).

\subsection{$N > 4$}

It seems quite natural to conjecture, based on the results of this section,
that, in general, factors of the form $(N +1 -N q)$ will appear in parallel
results for $N > 4$. 
We remark, in this regard, that for the case $N=2,m=4$, the denominators
of the three distinct eigenvalues of the $16 \times 16$
mean density matrix are all proportional to $(5 -2 q)(3 -2 q)$
(cf. (\ref{eigen5252}) - (\ref{eigen5353})).
\section{CONCLUDING REMARKS}
Of course, it would be of interest to obtain analogs of the
results
presented above for additional values of $N$ and/or $m$, and to explore areas
of further possible application (cf. \cite{pauncz1,pauncz2}).
The relation of the orthonormal bases of $N^{m}$-dimensional Hilbert space
that we have reported, to other bases, such as the Bell ($N=2,m=2$) and
Greenberger-Horne-Zeilinger ($N=2,m > 2$)
 ones, used in the processing of quantum information
\cite{zeilinger1,zeilinger2,gisin},
 merits investigation.  (These latter bases correspond to {\it maximally
entangled} states.)
In this regard, we make the simple observation that in the $N=2,m=2$
case, the eigenvectors found by Krattenthaler and Slater \cite{kratt}
 separate into a 
triplet ($M_{2,0} =3$) and singlet ($M_{2,1}=1$),
as with the Bell basis.
The triplet (and, of course, the singlet) is known to be irreducible
\cite[p. 103]{greiner}. We have, however, not formally established the
irreducibility of the eigenspaces for the $N > 3$ cases reported above
(secs.~\ref{N3case}, \ref{N6case}, \ref{N144one} and \ref{secN4}).

It would be desirable to better understand the significance of the process 
(selection rule) by
which we have annihilated those off-diagonal entries in the 
averaged density matrices (based on tensor
products incorporating $3 \times 3$ density matrices)
 which are equal to a rational number divided by $ \pi$ 
(leaving undisturbed those off-diagonal entries which are 
simply rational numbers)
 and thereby obtained eigenvalues, the multiplicities of which
correspond to the (permuationally-symmetrized) multiplets.
In this regard, it would be of interest to determine the relations of the
approach taken here to the alternative
one  of Katriel, Paldus and Pauncz \cite{katriel} (from whose article
 we have
excerpted certain passages at the end of sec.~\ref{explicitspin}).

We should also note for the benefit of those who might desire to further
pursue the lines of investigation followed here that our results 
 have, in large part,
been based on MATHEMATICA \cite{wolfram} computations,
 in particular, multiple integrations
over multivariate polynomials in trigonometric functions (sines
and cosines of single angles). However, since MATHEMATICA
would not directly integrate the 
cumbersome expressions (at least in an acceptable
amount of time), we found it necessary at each integration stage
to reduce the problem to its simplest possible
 (univariate) form, saving auxiliarly the unused
information for the next integration step. In fact, it appeared
necessary at several points to pursue even more subtle
strategies in order to reduce the computational (space and time)
 burden. In any case,
 we have attempted to present the most extensive
analyses within our capabilities of achieving.
We should also observe that we have not investigated, to any major extent,
the possible role for {\it numerical}, as opposed to exact or
symbolic,  integration methods.

\acknowledgments

I would like to express appreciation to the Institute for Theoretical
Physics for computational support in this research, C. Krattenthaler and
T. Brun
for certain technical advice, and D. Eardley for a discussion concerning
\cite{zycz}.

\listoffigures

\end{document}